# Positively Charged Additives Facilitate Incorporation in Inorganic Single Crystals

*Ouassef Nahi,[1]\* Alexander Broad,[2] Alexander N. Kulak,[1] Helen M. Freeman,[3] Shuheng Zhang,[1] Thomas D. Turner,[1] Lucien Roach,[4] Robert Darkins,[2] Ian J. Ford,[2] and Fiona C. Meldrum[1]\**

[1] School of Chemistry, University of Leeds, Woodhouse Lane, Leeds, LS2 9JT, UK

[2] Department of Physics and Astronomy, University College London, Gower Street, London WC1E 6BT, UK

[3] School of Civil Engineering, University of Leeds, Woodhouse Lane, Leeds, LS2 9JT, UK

[4] CNRS, Univ. Bordeaux, Bordeaux INP, ICMCB, UMR 5026, 33600 Pessac, FRANCE

**ABSTRACT**

Incorporation of guest additives within inorganic single crystals offers a unique strategy for creating nanocomposites with tailored properties. While anionic additives have been widely used to control the properties of crystals, their effective incorporation remains a key challenge. Here, we show that cationic additives are an excellent alterative for the synthesis of nanocomposites, where they are shown to deliver exceptional levels of incorporation of up to 70 wt% of positively charged amino acids, polymer particles, gold nanoparticles, and silver nanoclusters within inorganic single crystals. This high additive loading endows the nanocomposites with new functional properties including plasmon coupling, bright fluorescence, and surface-enhanced Raman scattering (SERS). Cationic additives are also shown to outperform their acidic counterparts, where they are highly active in a wider range of crystal systems, owing to their outstanding colloidal stability in the crystallization media and strong affinity for the crystal surfaces. This work demonstrates that although often overlooked, cationic additives can make valuable crystallization additives to create composite materials with tailored composition-structure-property relationships. This versatile and straightforward approach advances the field of single-crystal composites and provides exciting prospects for the design and fabrication of new hybrid materials with tunable functional properties.

**INTRODUCTION**

Incorporation of additives within single crystals holds enormous potential for the design and fabrication of composite materials with superior optical, electrical, mechanical and catalytic properties.[1-7] The single crystal matrix ensures that the additives are completely isolated from the surrounding environment, and the hybrid crystals possess structures that could not be replicated by simple mixing of the individual components. Incorporation of amino acids in calcite ($CaCO_3$) enhances the hardness of the hybrid crystals,[2] while improved stability and increase in band gap is achieved for organometal halide perovskites and ZnO crystals.[8-10] The occlusion of functional nanoparticles is particularly attractive, as their properties can be pre-defined using well-established synthetic methods. For example, incorporation of metal oxide nanoparticles[11-13] in $KH_2PO_4$ and $K_2SO_4$ has created single crystals for optoelectronic and photonic devices, while calcite containing polymer nano-objects[14] and magnetic nanoparticles[15] exhibit enhanced physical properties. Incorporation of gold nanoparticles[3] in ZnO crystals delivered nanocomposites with tunable band gaps and superior photocatalytic activity.

Despite these achievements, the occlusion of high concentrations of additives in single crystals remains challenging.[16-18] Indeed, only a few weight percent of quantum dots, Au and $Fe_3O_4$ nanoparticles were incorporated within calcite[19, 20], $Cu_2O$[21, 22], NaCl and borax[7] single crystals using gel-trapping, confinement-based and evaporation strategies, respectively. Nevertheless, it is possible to achieve homogeneous incorporation using simple co-precipitation,[3, 23-25] provided that an appropriate balance is achieved between additive binding and the rate of crystal growth. The additives must bind sufficiently strongly to the crystal surface such that they are engulfed by the subsequent steps, but not so strongly that they inhibit crystal growth.[26-28]

This strategy has been elegantly demonstrated for $CaCO_3$, where a broad range of organic (*e.g.,* polymer nanoparticles and protein nanogels)[14, 23, 29, 30] and inorganic (*e.g.,* carbon dots, Au and $Fe_3O_4$) nanoparticles[15, 24, 31] have been incorporated within calcite single crystals. Occlusion was achieved by functionalizing the nanoparticles with anionic polyelectrolytes rich in carboxylate,[30, 32, 33] sulfate,[34] sulfonate[35] or phosphate[36] functional groups such that they bind strongly to the crystal surfaces. However, this could only be carried out at low solution supersaturations and particle concentrations due to the colloidal instability of these anionic particles in the presence of calcium ions.[26] This, in turn, limits the scalability and thus the practical applications of this approach. Nanoparticles are therefore required that have a high affinity for the crystal surface, and that retain their colloidal stability at high calcium ion concentrations.[24]

Here, we show that these challenges can be overcome by functionalizing the particulate additives with cationic polyelectrolytes. Exceptional levels of incorporation (up to 70 wt%) of latex particles, gold nanoparticles and silver nanoclusters are achieved within calcite and alternative carbonate, sulfate, and oxide single crystals. Insight into the mechanism of occlusion and the structure/composition relationships of the nanocomposites is obtained from molecular dynamics simulations and synchrotron high-resolution powder X-ray diffraction. We also demonstrate that such high nanoparticle loadings lead to nanocomposites with new properties including bright fluorescence and surface-enhanced Raman scattering (SERS).

While anionic molecules have been the predominant choice of crystallization additive for inorganic compounds, this work shows that cationic species can outperform their anionic counterparts in some applications.

**RESULTS**

**Incorporation of small basic molecules in calcite**

Initial experiments investigated the precipitation of calcium carbonate in the presence of the basic amino acids L-arginine (Arg) and L-lysine (Lys). Precipitation was carried out using the ammonia diffusion method (pH ≈ 9),[37] where this yielded calcite under all of the conditions investigated ([$Ca^{2+}$] = 10 mM – 20 mM) and [amino acid] = 50 mM – 400 mM) (Figure 1 and Figures S1 and S2). Arg had no effect on the crystal morphologies, while Lys yielded rough crystals that were highly elongated along their $c$-axes. This morphology change is indicative of the preferential interaction of Lys with the acute over the obtuse steps of calcite.[26, 38] Thermogravimetric analyses (TGA) of crystals grown at [$Ca^{2+}$] = 20 mM and [amino acid] = 400 mM revealed that only ≈ 0.8 mol% of Arg incorporated in calcite, while up to ≈ 4 mol% was recorded for Lys (Figure 1e).

Molecular simulations were performed to elucidate the contrasting effects of Lys and Arg. The binding free energies for molecules representing the Lys and Arg side chains were computed at the terrace, acute step and acute kink sites of calcite (Figure 2a). Both molecules were modelled in a protonated state (p$K_a$ side chain of Lys = 10.6 and Arg = 12) reflecting the basicity of the experiments. The primary (-$NH_3$) amine in the Lys side chain interacted strongly with the carbonate anions in all adsorption sites; the adsorption free energy increased roughly in proportion to the number of carbonate anions available — one for a terrace, two for a step and three for a kink (Figure 2b). In contrast, the binding free energy of the Arg side chain was weaker at all adsorption sites, especially the step and kink sites. The stereochemistry and charge density of the single amine in the Lys side chain therefore provides stronger binding to calcite than the three amines in the Arg side chain, consistent with its enhanced occlusion and morphological effects.

**Incorporation of cationic polymer nanoparticles in calcite**

The incorporation of larger cationic particulate additives was then assessed. Poly(methyl methacrylate) (PMMA) nanoparticles that were surface-functionalized with branched poly(ethyleneimine) (PEI) were synthesized.[39] Scanning electron microscopy (SEM) and transmission electron microscopy (TEM) showed that the dry particles were highly monodisperse 150 nm spheres (Figure 3a-b) with a well-defined core/shell structure and a ≈ 15 nm PEI coating (Figure 3b). Dynamic light scattering (DLS) of the particles in water confirmed their uniformity (polydispersity index (PDI) = 0.05) and a mean hydrodynamic diameter of 200 nm (Figure S3), while zeta-potential measurements showed that they were positively charged at pH = 9 used for $CaCO_3$ precipitation (+45.7 mV).[40]

Calcite crystals precipitated in the presence of the nanoparticles were elongated along the *c*-axes and capped by {104} faces. Greater morphological effects were achieved at higher nanoparticle concentrations and supersaturations (Figure 3c-j) as the latter increases the step density, and therefore the growth rate, of the {104} faces, reducing their expression in the morphology. Given that the steps on the crystal surface are orders of magnitude smaller than the diameters of the nanoparticles, the specificity to acute over obtuse steps can be attributed to the binding of the cationic PEI chains on the surface of the polymer particles to these nanoscale features. Focused ion beam (FIB) milling or fractured crystals revealed uniform incorporation of the nanoparticles (Figure 3c-j), and TGA showed that the crystals comprise up to 33 wt% (57 vol%) nanoparticles (Figure S4).

**Comparison with low charge and anionic polymer nanoparticles**

The incorporation of the cationic PMMA-PEI nanoparticles was then compared with 200 nm low-charge and anionic nanoparticles. Non-functionalized poly(styrene) nanoparticles failed to incorporate or alter the crystal morphology (Figure 4d-f and Figure S4), while carboxyl-functionalized nanoparticles induced the same pattern of morphological changes as the cationic nanoparticles (Figure 5). Although the anionic particles incorporated better than cationic particles at low calcium ion concentrations ([$Ca^{2+}$] = 1.5 mM and [particles] = 0.1 wt%), a small increase to [$Ca^{2+}$] = 2.5 mM resulted in a reversal of their activities. Homogeneous incorporation of cationic particles occurred at [$Ca^{2+}$] = 2.5 mM and [particles] = 0.1 wt%, while 0.5 wt% of the anionic particles was required to give equivalent occlusion under the same conditions. This is attributed to the excellent colloidal stability of the cationic particles in the crystallization solution, which ensures that they remain well-dispersed in solutions of [$Ca^{2+}$] ≤ 50 mM (Figure S5). In contrast, the carboxyl-functionalized nanoparticles aggregated at much lower concentration of [$Ca^{2+}$] > 5 mM (Figure S6). Strong complexation of the carboxylated nanoparticles with $Ca^{2+}$ ions in solution also screens the negative charges and hampers their binding to the crystal surface at higher calcium concentrations.

**Incorporation of cationic metal nanoparticles in calcite**

This approach was then extended to create functional nanocomposite crystals using cationic metal nanoparticles.[24, 41, 42] Gold nanoparticles (AuNPs) functionalized with branched PEI (Au/PEI) were synthesized by complexing hydrochloroauric acid with branched PEI ($M_W$ = 1,200 g mol$^{-1}$), and then reducing with sodium borohydride. The average AuNPs diameter was 5 nm (± 1nm) as determined by TEM (Figure 6a). The AuNPs were highly stable in the mineralization solution, where DLS revealed only a minor increase in the hydrodynamic diameter from ≈ 6.5 nm to 8 nm as [$Ca^{2+}$] was increased from 0 mM to 30 mM (Figure S7). This is due to the slight expansion of the PEI polymers capping the AuNPs. Zeta-potential analysis confirmed that the nanoparticles were positively charged at pH = 9 (Figure S7), and the invariance of the surface plasmon resonance band at $\lambda$ = 520 nm demonstrated their stability in aqueous solutions containing [$Ca^{2+}$] ≤ 50 mM (Figure S8). This is a significant advantage over occlusion systems based on nanoparticles functionalized with anionic polymers, where these readily aggregate in the presence of metal ions.[24, 35]

Pink calcite crystals with non-uniform colors characteristic of intra-sectoral zoning formed on co-precipitation at very low [Au/PEI] = 0.01 wt% (Figure 6b), while higher [Au/PEI] = 0.1 wt% yielded elongated, dark red/black crystals (Figure 6c). The AuNPs were uniformly distributed throughout the host crystal (Figure 6d-f, Figure S9), and analysis of bulk samples using atomic absorption spectroscopy (AAS) and TGA showed that they comprise ≈ 43 wt% Au and ≈ 28 wt% PEI (Figure S10), and thus a remarkable 70 wt% of foreign material in calcite. While one may intuitively expect a disruption of the structure of the host crystal, both high-resolution TEM (HRTEM) (Figure 6g) and selected area electron diffraction (SAED) (Figure 6h-i) of a thin section cut from a nanocomposite crystal demonstrates that the single crystal structure of the calcite host is preserved.

This approach was also successfully applied to yet smaller silver nanoparticles (Ag/PEI) prepared by slow reduction of a solution of silver nitrate and branched PEI ($M_W$ = 10,000 g $mol^{-1}$) with L-ascorbic acid. This gave Ag nanoclusters with < 2nm Ag cores (Figure 7a and Figure S12) and ≈ 4 nm hydrodynamic diameter (Figure S11), which exhibit bright fluorescence under UV irradiation ($\lambda = 365$ nm) (inset Figure 7d).[43] Comparison of calcite growth in the presence of [nanoparticles] = 0.01 wt% showed that the Ag/PEI and Au/PEI nanoparticles induced the same pattern of morphological changes (Figure 7b), but that the Ag/PEI clusters were more effective. This is attributed to a higher number density of the smaller Ag/PEI nanoclusters. These crystals again exhibited zoning, where the nanoclusters concentrated in the rough equatorial region of calcite (Figure 7c).

Higher [Ag/PEI] = 0.1 wt% resulted in their uniform incorporation (Figure 7d and Figure S13) and endowed the host crystal with bright fluorescence under UV light (Figure 7e). Analysis of bulk samples using inductively coupled plasma-optical emission spectroscopy (ICP-OES) revealed Ag incorporation levels of 37.5 wt%, and TGA showed that the crystals comprised 26.3 wt% PEI (Figure S10). This again demonstrates that exceptional amounts – 64 wt% – of Ag clusters can be incorporated in calcite while preserving the single crystal structure (Figure 6h).

**Microstructure of the nanocomposite crystals**

The influence of the occluded PMMA-PEI, Au/PEI and Ag/PEI nanoparticles on the structure of the calcite host crystal was investigated using synchrotron high-resolution powder X-ray diffraction (HRPXRD), where data were compared with pure calcite and calcite containing PEI (Figure 8 and Figure S15). Lattice distortions, microstrain fluctuations and coherence lengths were measured for the (012), (104), (006) and (110) reflections, giving information about the additive/crystal interactions along different crystallographic directions.

PMMA-PEI nanoparticles only had minor effects on the calcite structure, even at incorporation levels of 57 vol%, causing a decrease in coherence lengths from 650 nm in pure calcite to 540 nm, and small lattice distortions of $\Delta a/a = 0.004$ % and $\Delta c/c = 0.01$ %. This is consistent with previous analysis of polymer nano-objects/calcite composites.[30] By comparison, the Bragg peaks of calcite containing ≈ 18 wt% PEI were significantly broadened and shifted to larger d-spacings. This sample also showed smaller coherence lengths of 470 nm and 330 nm perpendicular to the (110) and (006) planes, and lattice expansions of $\Delta a/a = 0.013$ % and $\Delta c/c = 0.095$ %. The greater lattice distortions along the *c*-axis than the *a*-axis (Figure 7f) are consistent with the elastic anisotropy of calcite.[2]

Incorporation of 64 wt% Ag/PEI and 70 wt% Au/PEI nanoparticles yielded significant broadenings and shifts of the diffraction peaks toward larger d-spacings. Coherence lengths were ≈ 350 nm and ≈ 210 nm perpendicular to the (110) and (006) planes for both nanocomposites, and lattice expansions reaching up to $\Delta c/c = 0.135$ % and $\Delta a/a = 0.03$ % were recorded. These values are high for a brittle ceramic and are comparable to those measured in calcite biominerals occluding biomolecules.[44] The strong influence of these small particles can be attributed to the large additive/host interfacial area, where the specific surface areas of the ≈ 6.5 nm Au/PEI particles and ≈ 4 nm Ag/PEI are ≈ 1,000 times and ≈ 2,500 times larger than that of the 200 nm latex particles, respectively. That the Au/PEI nanoparticles induced slightly greater lattice expansions and microstrains may derive from the composition of the nanoparticles, where the Ag/PEI nanoclusters comprise a higher percentage of polymer than the Au/PEI nanoparticles. This suggests that the metal nanoparticle core and the polymer coating have different effects on the crystal lattice.

**Incorporation in alternative host crystals**

The versatility of this approach was then demonstrated by extending it to manganese carbonate ($MnCO_3$, rhodochrosite), strontium sulfate ($SrSO_4$, celestine), calcium sulfate dihydrate (gypsum) and zinc oxide (ZnO, zincite) (Figure 9 and Figures S16–S19). Incorporation of AuNPs within $MnCO_3$, $SrSO_4$ and $CaSO_4 \cdot 2H_2O$ was achieved by mixing 0.1 wt% Au/PEI with the respective crystallization solutions. Au/ZnO nanocomposites were prepared under mild hydrothermal conditions in which 1 wt% Au/PEI was added to an aqueous solution of zinc nitrate hexahydrate and HMTA, and the reaction mixture was heated at 90°C for 90 min. High levels of occlusion were achieved in all crystals. The nanoparticles were uniformly distributed throughout ZnO, $MnCO_3$ and $SrSO_4$, where the ZnO contained ≈ 22.5 wt% Au/PEI, and the $MnCO_3$ and $SrSO_4$ over ≈ 65 wt%. Gypsum displayed pronounced sectoral zoning (Figure 9g) that is common for this crystal and derives from preferential binding of the AuNPs to the (011) planes.[45] Increasing the amount of Au/PEI in solution to 0.25 wt% enabled uniform incorporation of the AuNPs in gypsum (≈ 47 wt%) (Figure S19).

**Optical properties of the nanocomposite crystals**

The experimental extinction spectra of the Au/calcite crystals show that the Au plasmon peak is broadened and shifted to ≈ 557 nm as compared with ≈ 520 nm for AuNPs dispersed in aqueous solution (Figure 10b). This is consistent with finite element simulations of a pair of 5 nm AuNPs embedded in calcite, which predict a red-shift in the resonance peak from 520 nm to 560 nm due to the high refractive index of the calcite (n = 1.66) when the AuNPs are separated by 4 nm, and ultimately the development of two separate peaks as the inter-particle distance is decreased (Figure 10c).

The nanocomposite crystals also display surface-enhanced Raman scattering (SERS),[46] where this effect arises from the close proximity of the AuNPs (down to ≈ 1 nm) in the host crystals (Figure 10a), creating intense localized electromagnetic fields in the regions between the AuNPs.[46, 47] The main Raman peaks (*i.e.,* 1085 cm$^{-1}$ for calcite and $MnCO_3$, 1002 cm$^{-1}$ for SrSO$_4$, 1008 cm$^{-1}$ for gypsum, and 437 cm$^{-1}$ for ZnO) of the composites are significantly more intense than those of the pure crystals, with enhancement factors of 2 and 4.5 for ZnO and gypsum respectively, and ≈ 6 for calcite, MnCO$_3$ and SrSO$_4$ (Figure 10a and Figures S16 – S19). An inverse relationship exists between the SERS signal enhancement and the interparticle distance between adjacent AuNPs, such that the values recorded scale with the levels of Au incorporation within the various host crystals.[46, 48]

The experimental SERS data were rationalized by the simulations, where the increase in the Raman signals, $^{em}G_{SERS}$, is proportional to the fourth power of the local field enhancement, $|E|/|E_0|$, where $E$ is the local electric field and $E_0$ is the incident electric field.[49] The simulations predict field enhancements $^{em}G_{SERS} \approx 10^3$ at the midpoint between two AuNPs separated by 4 nm, where the induced field is inversely proportional to the third power of the distance from the particle surface.[50] Hotspots on the AuNPs surfaces with $^{em}G_{SERS}$ of $10^5$ - $10^8$ are also predicted for particle separations of 1 to 4 nm (Figure 10d-g), but these are unlikely to be fully accessible by the calcite crystal due to the PEI coating on the AuNPs. The field enhancement observed in the nanocomposite will also be reduced due to the presence of multiple AuNPs in close proximity, which will dilute the peak field enhancement and lower $^{em}G_{SERS}$ (Figures S20 and S21).

**DISCUSSION**

Incorporation of soluble additives within inorganic single crystals holds enormous potential for the synthesis of hybrid materials with advanced properties. However, the creation of such materials is far from trivial, as the simple addition of particles to a crystallization solution typically leads to low levels of occlusion.[16, 17] This has been addressed by functionalizing particles with block copolymers that modify the interaction of the particles with the crystal surface to enhance their occlusion, and that ensure their colloidal stability in the crystallization solution.[26, 35]

Anionic polyelectrolytes have been used extensively to functionalize particles and drive incorporation.[18, 32-36] This selection is made based on the traditional choice of anionic additives for controlling $CaCO_3$ precipitation, where they are often highly active in modifying morphologies and retarding nucleation.[26, 35] Highly acidic macromolecules are also characteristic of calcium carbonate biomineralization, and are thus often employed in bio-inspired mineralization.[14, 23, 30] However, solution conditions are restricted to low supersaturations and particle concentrations as strong electrostatic interactions between the calcium ions and anionic moieties on the surface of the particles cause particle aggregation. This significantly limits the particle loading and yield of these nanocomposites.

Many studies have been performed to determine "design rules" that govern the occlusion of nanoparticles functionalized with polymer chains. As a general rule, short anionic polymer chains often lead to occlusion in the outer surface of the calcite crystals only,[17, 26, 35] while longer chains deliver incorporation throughout the crystals.[14, 26, 30] This behavior has been attributed to the higher conformational freedom of the longer chains, which enhances binding of the particles to the surface of the mineral.[26] Exceptions are noted, however, where longer sulfate chains were observed to give less incorporation than their shorter counterparts,[34] even at low $[Ca^{2+}]$ = 1.5 mM.

The complexity of occluding anionic particles was further investigated by designing block copolymer micelles in which the ratio of carboxylate- to hydroxyl-functionalized chains could be systematically varied.[26] Nanoparticle incorporation in calcite did not directly scale with the carboxylate content of the steric stabilizers, where nanoparticles comprising 1:1 carboxylate/ hydroxyl groups were occluded at higher levels than those with carboxylate chains only. In this scenario, the hydroxyl moieties were considered to reduce ionic cross-linking between adjacent carboxylates chains, providing higher colloidal stability and more efficient binding to the crystal surface. These studies show that the incorporation of anionic particles in calcite is a complex process relying on trial and error to achieve success.

Cationic polyelectrolytes, in contrast, offer a straightforward and robust means of synthesizing hybrid crystals, where the weak interactions between the $Ca^{2+}$ ions and cationic polymers ensures excellent colloidal stability of the nanoparticles. High concentrations of nanoparticles can therefore be employed. Although the cationic additives bind less strongly to the crystal surface than their anionic counterparts,[51] access to much higher particle concentrations and supersaturations delivers exceptional occlusion levels: 57 vol% polymer nanoparticles, 64 wt% silver nanoclusters and 70 wt% gold nanoparticles. This is achieved for a broad range of supersaturations and without compromising the single crystallinity of calcite, which enables scale-up. The polyamines employed were all commercially available, avoiding the need for the synthesis of bespoke polymers.

These results also contribute to the growing recognition that cationic additives can make valuable crystallization additives. While they have been largely neglected in the control of calcium carbonate and other inorganic crystals, some notable exceptions exist such as the formation of fibers and thin films of $CaCO_3$ in the presence of poly(allylamine hydrochloride) (PAH).[52, 53] Organic molecules bearing amine moieties,[54] basic polypeptides[55] and cationic polyelectrolytes[40] can additionally enable polymorph selection, delivering metastable vaterite and aragonite over calcite (the thermodynamically stable phase).[54, 56, 57] Positively charged molecules may also play important roles in biogenic systems, where analysis of the structures of mollusk shells have identified biomacromolecules rich in histidine, arginine and lysine residues.[58-60] Our demonstration that polyamines can drive the occlusion of particles to exceptionally high levels therefore further extends the scope of these molecules as crystallization additives.

## CONCLUSIONS

While cationic additives are often overlooked in favor of their anionic counterparts, growing evidence suggests that they can play pivotal roles in controlling the crystallization of inorganic compounds. This work demonstrates that functionalization of particulate additives with commercially-available cationic polyelectrolytes offers a robust and versatile means of fabricating functional nanocomposites of a wide range of inorganic single crystals including carbonates, sulfates and oxides. Exceptional levels of occlusion of organic and inorganic particulate additives were achieved, far exceeding those obtained with their acidic counterparts. This work therefore provides a significant step-change in methodology, where we envisage that SERS combined with the intrinsic plasmonic properties of metal nanoparticles and nanoclusters within inorganic single crystals could find immediate applications in a wide range of areas including catalysis, bioimaging, sensing and photothermal therapy.

## ASSOCIATED CONTENT

**Supporting Information**

Experimental details including the synthesis and characterization of the polymer and metal nanoparticles, their incorporation in calcite and alternative host single crystals, additional optical micrographs, DLS, electrophoretic, UV-Visible, Raman, HRPXRD, TGA, TEM-EDX, SEM-EDX and FIB-SEM analyses, and supplementary information about the MD simulations.

## AUTHOR INFORMATION

Corresponding Authors

Ouassef Nahi, School of Chemistry, University of Leeds, Woodhouse Lane, Leeds, LS2 9JT, UK.

Email: pmona@leeds.ac.uk

Fiona C. Meldrum, School of Chemistry, University of Leeds, Woodhouse Lane, Leeds, LS2 9JT, UK.

Email: f.meldrum@leeds.ac.uk

**Notes**

The authors declare no competing financial interest.

## ACKNOWLEDGEMENTS

We are grateful to the Engineering and Physical Sciences (EPSRC) for financial support for O.N. through the Centre for Doctoral Training in Complex Particulate Products and Processes (EP/L015285/1) and for funding *via* the Programme Grant (EP/R018820/1) which funds the Crystallization in the Real World consortium (FCM, HMF, RD and IJF) and project EP/T006331/1 (FCM and TDT). We thank the European Research Council (ERC) for funding the project DYNAMIN, grant agreement number 788968 (SZ and FCM) and the European Synchrotron Radiation Facility (ESRF – beamline ID22) for beamtime and the assistance provided by Dr. Ola G. Grendal. We are grateful to Dr. Zabeada Aslam, School of Chemical and Process Engineering (Leeds, UK), for the help with the TEM characterization.

We also thank Stephen Reid, School of Earth and Environment (Leeds, UK) for the help with the ICP-OES analyses.

**REFERENCES**

(1) Ning, Z.; Gong, X.; Comin, R.; Walters, G.; Fan, F.; Voznyy, O.; Yassitepe, E.; Buin, A.; Hoogland, S.; Sargent, E. H., Quantum-dot-in-perovskite solids. *Nature* **2015,** 523, 7560, 324-328.

(2) Kim, Y.-Y.; Carloni, J. D.; Demarchi, B.; Sparks, D.; Reid, D. G.; Kunitake, Miki E.; Tang, C. C.; Duer, M. J.; Freeman, C. L.; Pokroy, B.; Penkman, K.; Harding, J. H.; Estroff, L. A.; Baker, S. P.; Meldrum, F. C., Tuning hardness in calcite by incorporation of amino acids. *Nat. Mater.* **2016,** 15, 8, 903-910.

(3) Ning, Y.; Fielding, L. A.; Nutter, J.; Kulak, A. N.; Meldrum, F. C.; Armes, S. P., Spatially Controlled Occlusion of Polymer-Stabilized Gold Nanoparticles within ZnO. *Angew. Chem. Int. Ed.* **2019,** 58, 13, 4302-4307.

(4) Green, D. C.; Holden, M. A.; Levenstein, M. A.; Zhang, S.; Johnson, B. R. G.; Gala de Pablo, J.; Ward, A.; Botchway, S. W.; Meldrum, F. C., Controlling the fluorescence and room-temperature phosphorescence behaviour of carbon nanodots with inorganic crystalline nanocomposites. *Nat. Commun.* **2019,** 10, 1, 206-206.

(5) Kulak, A. N.; Yang, P.; Kim, Y.-Y.; Armes, S. P.; Meldrum, F. C., Colouring crystals with inorganic nanoparticles. *ChemComm* **2014,** 50, 1, 67-69.

(6) Muñoz-Espí, R.; Jeschke, G.; Lieberwirth, I.; Gómez, C. M.; Wegner, G., ZnO-latex hybrids obtained by polymer-controlled crystallization: a spectroscopic investigation. *J. Phys. Chem. B* **2007,** 111, 4, 697-707.

(7) Adam, M.; Erdem, T.; Stachowski, G. M.; Soran-Erdem, Z.; Lox, J. F.; Bauer, C.; Poppe, J.; Demir, H. V.; Gaponik, N.; Eychmüller, A., Implementation of High-Quality Warm-White Light-Emitting Diodes by a Model-Experimental Feedback Approach Using Quantum Dot-Salt Mixed Crystals. *ACS Appl. Mater. Interfaces* **2015,** 7, 41, 23364-71.

(8) Lang, A.; Polishchuk, I.; Seknazi, E.; Feldmann, J.; Katsman, A.; Pokroy, B., Bioinspired Molecular Bridging in a Hybrid Perovskite Leads to Enhanced Stability and Tunable Properties. *Adv. Funct. Mater.* **2020,** 30, 42, 2005136.

(9) Brif, A.; Bloch, L.; Pokroy, B., Bio-inspired engineering of a zinc oxide/amino acid composite: synchrotron microstructure study. *CrystEngComm* **2014,** 16, 16, 3268-3273.

(10) Brif, A.; Ankonina, G.; Drathen, C.; Pokroy, B., Bio-inspired band gap engineering of zinc oxide by intracrystalline incorporation of amino acids. *Adv Mater* **2014,** 26, 3, 477-81.

(11) Pritula, I. M.; Kosinova, A. V.; Kolybaeva, M. I.; Bezkrovnaya, O. N.; Grebenev, V. V.; Voloshin, A. E.; Vorontsov, D. A.; Sofronov, D. S.; Vovk, O. M.; Baumer, V. N., Some characteristic features of formation of composite material based on KDP single crystal with incorporated $Al_2O_3 \cdot nH_2O$ nanoparticles. *Cryst. Res. Technol.* **2014,** 49, 5, 345-352.

(12) Gayvoronsky, V. Y.; Kopylovsky, M. A.; Brodyn, M. S.; Pritula, I. M.; Kolybaeva, M. I.; Puzikov, V. M., Impact of incorporated anatase nanoparticles on the second harmonic generation in KDP single crystals. *Appl. Phys. Lett.* **2013,** 10, 3, 035401.

(13) Pritula, I.; Gayvoronsky, V.; Kolybaeva, M.; Puzikov, V.; Brodyn, M.; Tkachenko, V.; Kosinova, A.; Kopylovsky, M.; Tsurikov, V.; Bezkrovnaya, O., Effect of incorporation of titanium dioxide nanocrystals on bulk properties of KDP crystals. *Opt. Mater.* **2011,** 33, 4, 623-630.

(14) Kim, Y.-Y.; Ribeiro, L.; Maillot, F.; Ward, O.; Eichhorn, S. J.; Meldrum, F. C., Bio-Inspired Synthesis and Mechanical Properties of Calcite–Polymer Particle Composites. *Adv. Mater.* **2010,** 22, 18, 2082-2086.

(15) Kulak, A. N.; Semsarilar, M.; Kim, Y.-Y.; Ihli, J.; Fielding, L. A.; Cespedes, O.; Armes, S. P.; Meldrum, F. C., One-pot synthesis of an inorganic heterostructure: uniform occlusion of magnetite nanoparticles within calcite single crystals. *Chem. Sci.* **2014,** 5, 2, 738-743.

(16)     Borukhin, S.; Bloch, L.; Radlauer, T.; Hill, A. H.; Fitch, A. N.; Pokroy, B., Screening the Incorporation of Amino Acids into an Inorganic Crystalline Host: the Case of Calcite. *Adv. Funct. Mater.* **2012,** 22, 20, 4216-4224.

(17)     Lu, C.; Qi, L.; Cong, H.; Wang, X.; Yang, J.; Yang, L.; Zhang, D.; Ma, J.; Cao, W., Synthesis of Calcite Single Crystals with Porous Surface by Templating of Polymer Latex Particles. *Chem. Mater.* **2005,** 17, 20, 5218-5224.

(18)     Hanisch, A.; Yang, P.; Kulak, A. N.; Fielding, L. A.; Meldrum, F. C.; Armes, S. P., Phosphonic Acid-Functionalized Diblock Copolymer Nano-Objects via Polymerization-Induced Self-Assembly: Synthesis, Characterization, and Occlusion into Calcite Crystals. *Macromolecules* **2016,** 49, 1, 192-204.

(19)     Liu, Y.; Zang, H.; Wang, L.; Fu, W.; Yuan, W.; Wu, J.; Jin, X.; Han, J.; Wu, C.; Wang, Y.; Xin, H. L.; Chen, H.; Li, H., Nanoparticles Incorporated inside Single-Crystals: Enhanced Fluorescent Properties. *Chem. Mater.* **2016,** 28, 20, 7537-7543.

(20)     Liu, Y.; Yuan, W.; Shi, Y.; Chen, X.; Wang, Y.; Chen, H.; Li, H., Functionalizing single crystals: incorporation of nanoparticles inside gel-grown calcite crystals. *Angew. Chem. Int. Ed.* **2014,** 53, 16, 4127-31.

(21)     DiCorato, A. E.; Asenath-Smith, E.; Kulak, A. N.; Meldrum, F. C.; Estroff, L. A., Cooperative Effects of Confinement and Surface Functionalization Enable the Formation of Au/Cu2O Metal–Semiconductor Heterostructures. *Cryst. Growth Des.* **2016,** 16, 12, 6804-6811.

(22)     Asenath-Smith, E.; Noble, J. M.; Hovden, R.; Uhl, A. M.; DiCorato, A.; Kim, Y.-Y.; Kulak, A. N.; Meldrum, F. C.; Kourkoutis, L. F.; Estroff, L. A., Physical Confinement Promoting Formation of Cu2O–Au Heterostructures with Au Nanoparticles Entrapped within Crystalline Cu2O Nanorods. *Chem. Mater.* **2017,** 29, 2, 555-563.

(23)     Kim, Y. Y.; Ganesan, K.; Yang, P.; Kulak, A. N.; Borukhin, S.; Pechook, S.; Ribeiro, L.; Kröger, R.; Eichhorn, S. J.; Armes, S. P.; Pokroy, B.; Meldrum, F. C., An artificial biomineral formed by incorporation of copolymer micelles in calcite crystals. *Nat. Mater.* **2011,** 10, 11, 890-6.

(24)     Kim, Y.-Y.; Darkins, R.; Broad, A.; Kulak, A. N.; Holden, M. A.; Nahi, O.; Armes, S. P.; Tang, C. C.; Thompson, R. F.; Marin, F.; Duffy, D. M.; Meldrum, F. C., Hydroxyl-rich macromolecules enable the bio-inspired synthesis of single crystal nanocomposites. *Nat. Commun.* **2019,** 10, 1, 5682.

(25)     Muñoz-Espí, R.; Qi, Y.; Lieberwirth, I.; Gómez, C. M.; Wegner, G., Surface-Functionalized Latex Particles as Controlling Agents for the Mineralization of Zinc Oxide in Aqueous Medium. *Chem. Eur. J.* **2006,** 12, 1, 118-129.

(26)     Kim, Y.-Y.; Fielding, L. A.; Kulak, A. N.; Nahi, O.; Mercer, W.; Jones, E. R.; Armes, S. P.; Meldrum, F. C., Influence of the Structure of Block Copolymer Nanoparticles on the Growth of Calcium Carbonate. *Chem. Mater.* **2018,** 30, 20, 7091-7099.

(27)     Rae Cho, K.; Kim, Y.-Y.; Yang, P.; Cai, W.; Pan, H.; Kulak, A. N.; Lau, J. L.; Kulshreshtha, P.; Armes, S. P.; Meldrum, F. C.; De Yoreo, J. J., Direct observation of mineral–organic composite formation reveals occlusion mechanism. *Nat. Commun.* **2016,** 7, 1, 10187.

(28)     Green, D. C.; Ihli, J.; Thornton, P. D.; Holden, M. A.; Marzec, B.; Kim, Y.-Y.; Kulak, A. N.; Levenstein, M. A.; Tang, C.; Lynch, C.; Webb, S. E. D.; Tynan, C. J.; Meldrum, F. C., 3D visualization of additive occlusion and tunable full-spectrum fluorescence in calcite. *Nat. Commun.* **2016,** 7, 1, 13524.

(29)     Nahi, O.; Kulak, A. N.; Kress, T.; Kim, Y.-Y.; Grendal, O. G.; Duer, M. J.; Cayre, O. J.; Meldrum, F. C., Incorporation of nanogels within calcite single crystals for the storage, protection and controlled release of active compounds. *Chem. Sci.* **2021,** 12, 28, 9839-9850.

(30) Kim, Y.-Y.; Semsarilar, M.; Carloni, J. D.; Cho, K. R.; Kulak, A. N.; Polishchuk, I.; Hendley IV, C. T.; Smeets, P. J. M.; Fielding, L. A.; Pokroy, B.; Tang, C. C.; Estroff, L. A.; Baker, S. P.; Armes, S. P.; Meldrum, F. C., Structure and Properties of Nanocomposites Formed by the Occlusion of Block Copolymer Worms and Vesicles Within Calcite Crystals. *Adv. Funct. Mater.* **2016,** 26, 9, 1382-1392.

(31) Green, D. C.; Holden, M. A.; Levenstein, M. A.; Zhang, S.; Johnson, B. R. G.; Gala de Pablo, J.; Ward, A.; Botchway, S. W.; Meldrum, F. C., Controlling the fluorescence and room-temperature phosphorescence behaviour of carbon nanodots with inorganic crystalline nanocomposites. *Nat. Commun.* **2019,** 10, 1, 206.

(32) Ning, Y.; Han, L.; Douverne, M.; Penfold, N. J. W.; Derry, M. J.; Meldrum, F. C.; Armes, S. P., What Dictates the Spatial Distribution of Nanoparticles within Calcite? *J. Am. Chem. Soc.* **2019,** 141, 6, 2481-2489.

(33) Ning, Y.; Whitaker, D. J.; Mable, C. J.; Derry, M. J.; Penfold, N. J. W.; Kulak, A. N.; Green, D. C.; Meldrum, F. C.; Armes, S. P., Anionic block copolymer vesicles act as Trojan horses to enable efficient occlusion of guest species into host calcite crystals. *Chem. Sci.* **2018,** 9, 44, 8396-8401.

(34) Ning, Y.; Fielding, L. A.; Ratcliffe, L. P. D.; Wang, Y.-W.; Meldrum, F. C.; Armes, S. P., Occlusion of Sulfate-Based Diblock Copolymer Nanoparticles within Calcite: Effect of Varying the Surface Density of Anionic Stabilizer Chains. *J. Am. Chem. Soc.* **2016,** 138, 36, 11734-11742.

(35) Ning, Y.; Han, L.; Derry, M. J.; Meldrum, F. C.; Armes, S. P., Model Anionic Block Copolymer Vesicles Provide Important Design Rules for Efficient Nanoparticle Occlusion within Calcite. *J. Am. Chem. Soc.* **2019,** 141, 6, 2557-2567.

(36) Douverne, M.; Ning, Y.; Tatani, A.; Meldrum, F. C.; Armes, S. P., How Many Phosphoric Acid Units Are Required to Ensure Uniform Occlusion of Sterically Stabilized Nanoparticles within Calcite? *Angew. Chem. Int. Ed.* **2019,** 58, 26, 8692-8697.

(37) Ihli, J.; Bots, P.; Kulak, A.; Benning, L. G.; Meldrum, F. C., Elucidating Mechanisms of Diffusion-Based Calcium Carbonate Synthesis Leads to Controlled Mesocrystal Formation. *Adv. Funct. Mater.* **2013,** 23, 15, 1965-1973.

(38) Orme, C. A.; Noy, A.; Wierzbicki, A.; McBride, M. T.; Grantham, M.; Teng, H. H.; Dove, P. M.; DeYoreo, J. J., Formation of chiral morphologies through selective binding of amino acids to calcite surface steps. *Nature* **2001,** 411, 6839, 775-779.

(39) Kook, J.-W.; Lee, J.; Hwang, K.; Park, I.; Kim, J., Synthesis and Characterization of Poly (Methyl Methacrylate)/Polyethylenimine Grafting Core-Shell Nanoparticles for CO2 Adsorption Using Soap-Free Emulsion Copolymerization. *AMPC* **2016,** 6, 7, 220-229.

(40) Kun Park, H.; Lee, I.; Kim, K., Controlled growth of calcium carbonate by poly(ethylenimine) at the air/water interface. *ChemComm* **2004**, 1, 24-25.

(41) Yeh, Y.-C.; Creran, B.; Rotello, V. M., Gold nanoparticles: preparation, properties, and applications in bionanotechnology. *Nanoscale* **2012,** 4, 6, 1871-1880.

(42) Huang, X.; El-Sayed, M. A., Gold nanoparticles: Optical properties and implementations in cancer diagnosis and photothermal therapy. *J. Adv. Res.* **2010,** 1, 1, 13-28.

(43) Santillán, J. M. J.; Muñetón Arboleda, D.; Muraca, D.; Schinca, D. C.; Scaffardi, L. B., Highly fluorescent few atoms silver nanoclusters with strong photocatalytic activity synthesized by ultrashort light pulses. *Sci. Rep.* **2020,** 10, 1, 8217.

(44) Pokroy, B.; Fitch, A.; Zolotoyabko, E., The Microstructure of Biogenic Calcite: A View by High-Resolution Synchrotron Powder Diffraction. *Adv. Mater.* **2006,** 18, 18, 2363-2368.

(45)     Aquilano, D.; Otálora, F.; Pastero, L.; García-Ruiz, J. M., Three study cases of growth morphology in minerals: Halite, calcite and gypsum. *Prog. Cryst. Growth Charact.* **2016,** 62, 2, 227-251.

(46)     Amendola, V.; Pilot, R.; Frasconi, M.; Maragò, O. M.; Iatì, M. A., Surface plasmon resonance in gold nanoparticles: a review. *J. Condens. Matter Phys.* **2017,** 29, 20, 203002.

(47)     Amendola, V.; Meneghetti, M., Exploring How to Increase the Brightness of Surface-Enhanced Raman Spectroscopy Nanolabels: The Effect of the Raman-Active Molecules and of the Label Size. *Adv. Funct. Mater.* **2012,** 22, 2, 353-360.

(48)     Zohar, N.; Chuntonov, L.; Haran, G., The simplest plasmonic molecules: Metal nanoparticle dimers and trimers. *J. Photochem. Photobiol.* **2014,** 21, 26-39.

(49)     Kneipp, K.; Wang, Y.; Kneipp, H.; Perelman, L. T.; Itzkan, I.; Dasari, R. R.; Feld, M. S., Single Molecule Detection Using Surface-Enhanced Raman Scattering (SERS). *Phys. Rev. Lett.* **1997,** 78, 9, 1667-1670.

(50)     Maier, S. A., *Plasmonics: Fundamentals and Applications.* New York: Springer: **2007**; Vol. 1.

(51)     Schenk, A. S.; Cantaert, B.; Kim, Y.-Y.; Li, Y.; Read, E. S.; Semsarilar, M.; Armes, S. P.; Meldrum, F. C., Systematic Study of the Effects of Polyamines on Calcium Carbonate Precipitation. *Chem. Mater.* **2014,** 26, 8, 2703-2711.

(52)     Cantaert, B.; Kim, Y.-Y.; Ludwig, H.; Nudelman, F.; Sommerdijk, N. A. J. M.; Meldrum, F. C., Think Positive: Phase Separation Enables a Positively Charged Additive to Induce Dramatic Changes in Calcium Carbonate Morphology. *Adv. Funct. Mater.* **2012,** 22, 5, 907-915.

(53)     Cantaert, B.; Verch, A.; Kim, Y.-Y.; Ludwig, H.; Paunov, V. N.; Kröger, R.; Meldrum, F. C., Formation and Structure of Calcium Carbonate Thin Films and Nanofibers Precipitated in the Presence of Poly(Allylamine Hydrochloride) and Magnesium Ions. *Chem. Mater.* **2013,** 25, 24, 4994-5003.

(54)    Xu, A.-W.; Antonietti, M.; Cölfen, H.; Fang, Y.-P., Uniform Hexagonal Plates of Vaterite CaCO3 Mesocrystals Formed by Biomimetic Mineralization. *Adv. Funct. Mater.* **2006,** 16, 7, 903-908.

(55)    Yao, Y.; Dong, W.; Zhu, S.; Yu, X.; Yan, D., Novel morphology of calcium carbonate controlled by poly(L-lysine). *Langmuir* **2009,** 25, 22, 13238-43.

(56)    Xu, A.-W.; Dong, W.-F.; Antonietti, M.; Cölfen, H., Polymorph Switching of Calcium Carbonate Crystals by Polymer-Controlled Crystallization. *Adv. Funct. Mater.* **2008,** 18, 8, 1307-1313.

(57)    Nassif, N.; Gehrke, N.; Pinna, N.; Shirshova, N.; Tauer, K.; Antonietti, M.; Cölfen, H., Synthesis of Stable Aragonite Superstructures by a Biomimetic Crystallization Pathway. *Angew. Chem. Int. Ed.* **2005,** 44, 37, 6004-6009.

(58)    Mann, K.; Siedler, F.; Treccani, L.; Heinemann, F.; Fritz, M., Perlinhibin, a Cysteine-, Histidine-, and Arginine-Rich Miniprotein from Abalone (Haliotis laevigata) Nacre, Inhibits In Vitro Calcium Carbonate Crystallization. *Biophys. J.* **2007,** 93, 4, 1246-1254.

(59)    Zhang, C.; Xie, L.; Huang, J.; Liu, X.; Zhang, R., A novel matrix protein family participating in the prismatic layer framework formation of pearl oyster, Pinctada fucata. *Biochem. Biophys. Res. Commun.* **2006,** 344, 3, 735-40.

(60)    Freeman, C. L.; Harding, J. H.; Quigley, D.; Rodger, P. M., Simulations of Ovocleidin-17 Binding to Calcite Surfaces and Its Implications for Eggshell Formation. *J. Phys. Chem.* **2011,** 115, 16, 8175-8183.

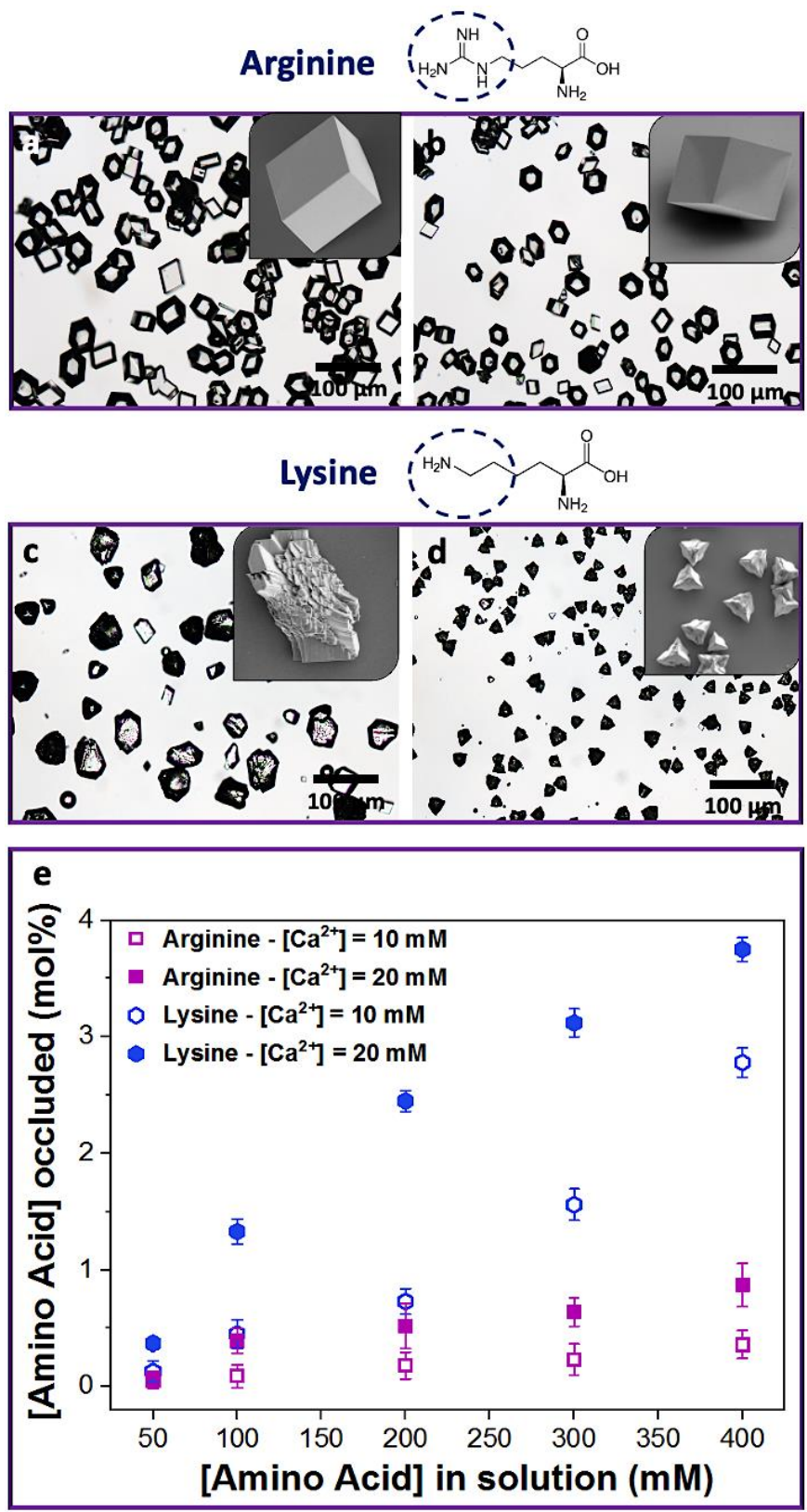

**Figure 1.** Optical micrographs of calcite precipitated in the presence of L-arginine (a-b) and L-lysine (c-d). Insets are the SEM images of the calcite crystals. (e) Graph showing the relationship between occlusion and the concentration of amino acid in solution.

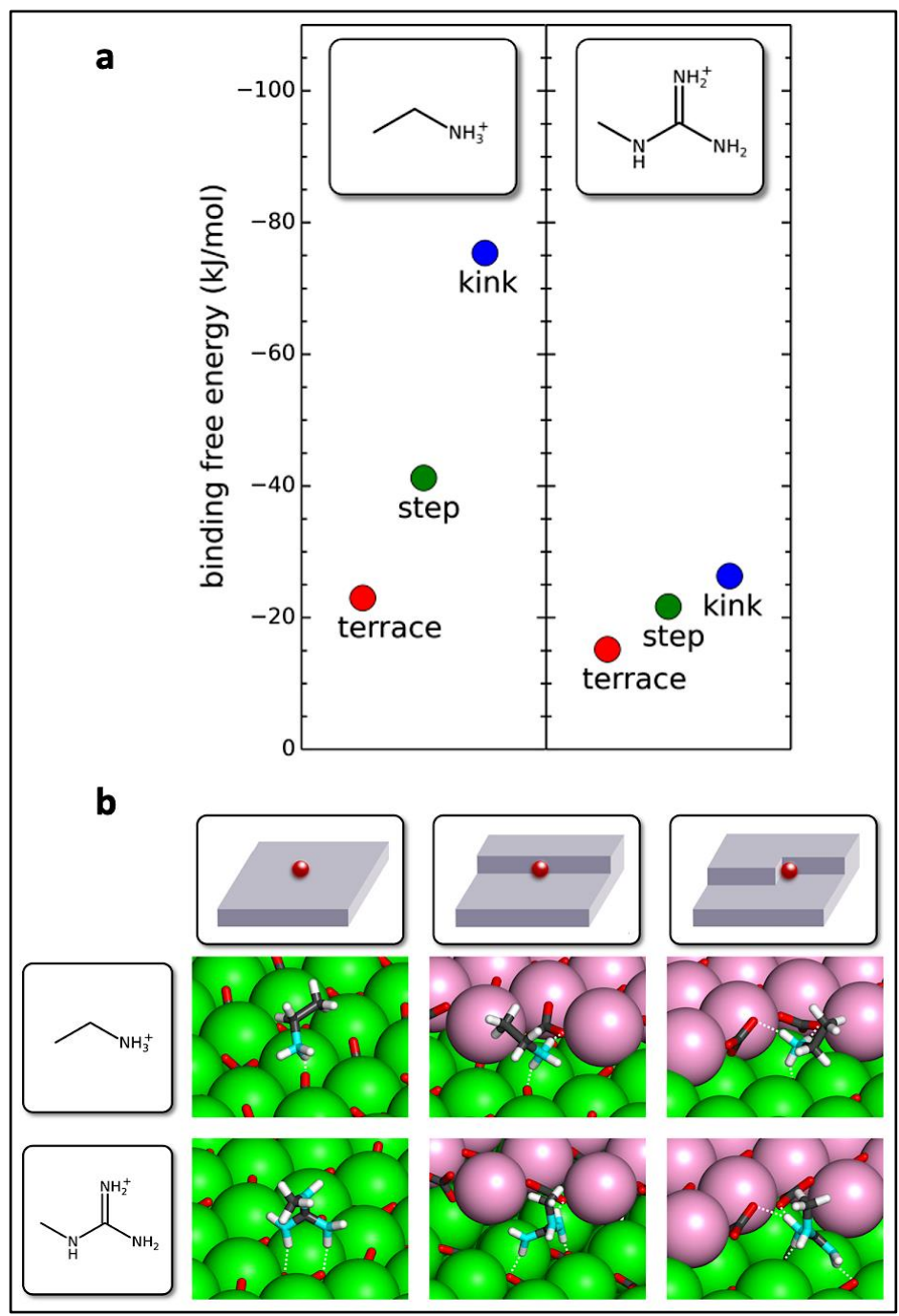

**Figure 2.** (a) Binding free energies for molecules representing the sides chains of L-lysine and L-arginine at the calcite terrace, acute step, and acute kink adsorption sites. (b) Snapshots from trajectories showing binding to the calcite terrace, step and kink sites. The white dashed lines indicate hydrogen bonds. Calcium is shown in green for the lower terrace and pink for the upper terrace. Carbon is shown in grey, oxygen in red, nitrogen in cyan and hydrogen in white. Water is excluded from all images for clarity.

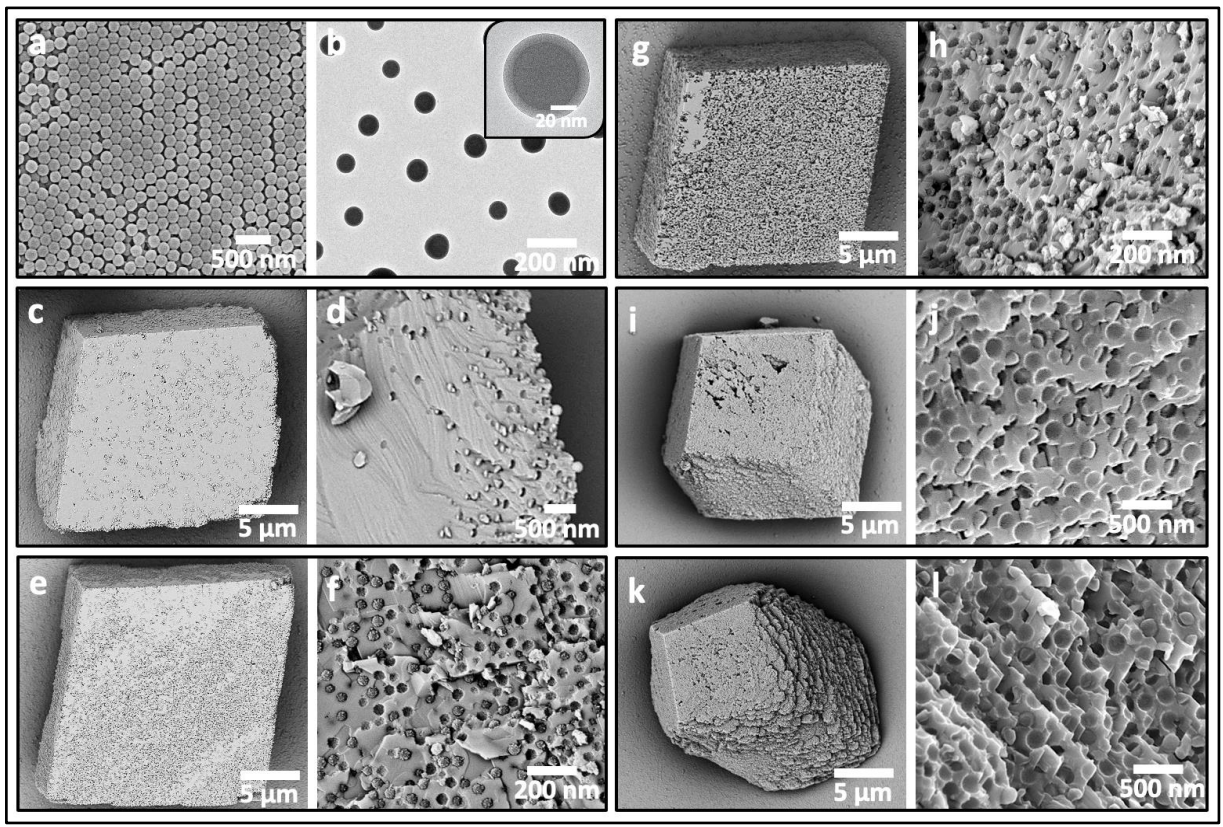

**Figure 3.** (a) SEM micrograph of the cationic PMMA-PEI nanoparticles. (b) TEM image showing that the nanoparticles are core-shell structures with a 15 nm PEI corona. (c-l) SEM images and cross-sections through PMMA-PEI/calcite composites precipitated in the presence of (c-h) 0.10 wt% and (i-l) 0.25wt% PMMA-PEI nanoparticles for (c-d) [$Ca^{2+}$] = 1.5 mM, (e-f and i-j) [$Ca^{2+}$] = 2.5 mM and (g-h and k-l) [$Ca^{2+}$] = 5 mM.

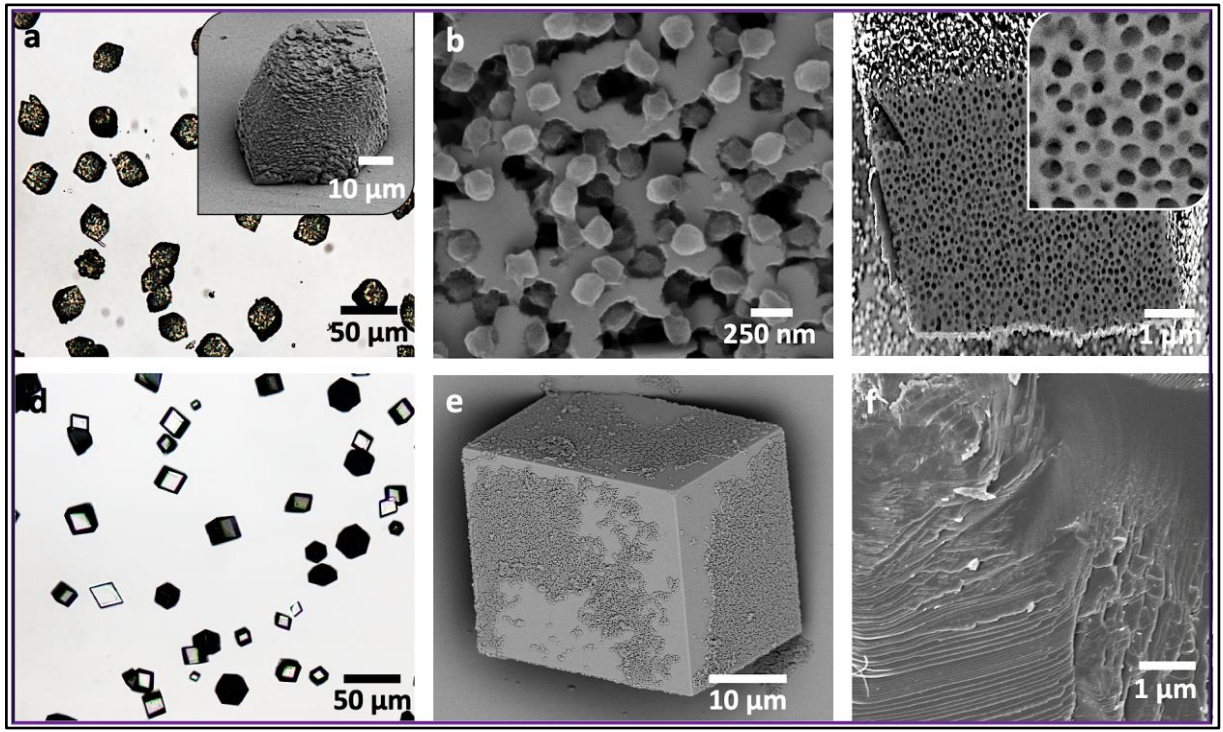

**Figure 4.** Optical (a and d) and SEM micrographs (inset in a, b-c and e-f) of calcite crystals precipitated in the presence of (a-c) PMMA-PEI nanoparticles and (d-f) non-functionalized PS nanoparticles at $[Ca^{2+}]$ = 10 mM and [nanoparticles] = 0.5 wt%. (b) Surface of PMMA-PEI/calcite nanocomposites showing embedded nanoparticles. (c) Cross-section showing uniform incorporation of the PMMA-PEI nanoparticles in calcite. (d-e) Images of unmodified calcite crystal precipitated in the presence of non-functionalized PS nanoparticles, and (f) cross-section showing no incorporation of nanoparticles in the bulk of the calcite crystals.

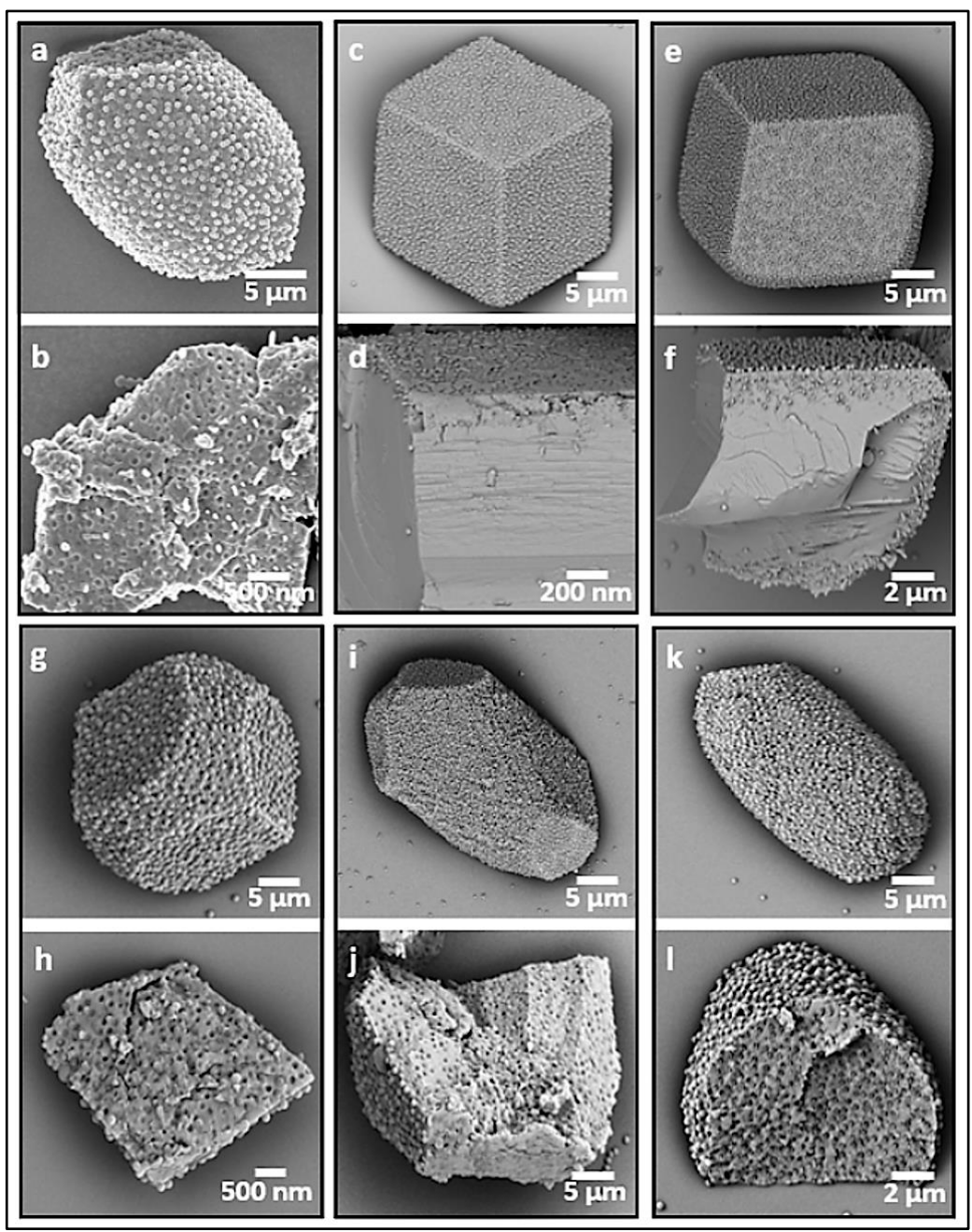

**Figure 5.** SEM micrographs of calcite crystals precipitated at [$Ca^{2+}$] = 1.5 mM (a-b) and [$Ca^{2+}$] = 2.5 mM (c-l) in the presence of carboxyl-functionalized nanoparticles at concentration (a-d) 0.1 wt%, (e-f) 0.25 wt%, (g-h) 0.5 wt%, (i-j) 1 wt%, (k-l) 2.5 wt%. SEM of fractured crystals showing that efficient incorporation is achieved at [$Ca^{2+}$] = 1.5 mM and [nanoparticles] = 0.1 wt% (b), whereas only surface-confined occlusion is achieved at [$Ca^{2+}$] = 2.5 mM and [nanoparticles] ≤ 0.25 wt% (d and f). Uniform incorporation occurs at [$Ca^{2+}$] = 2.5 mM and [nanoparticle] ≥ 0.5 wt%.

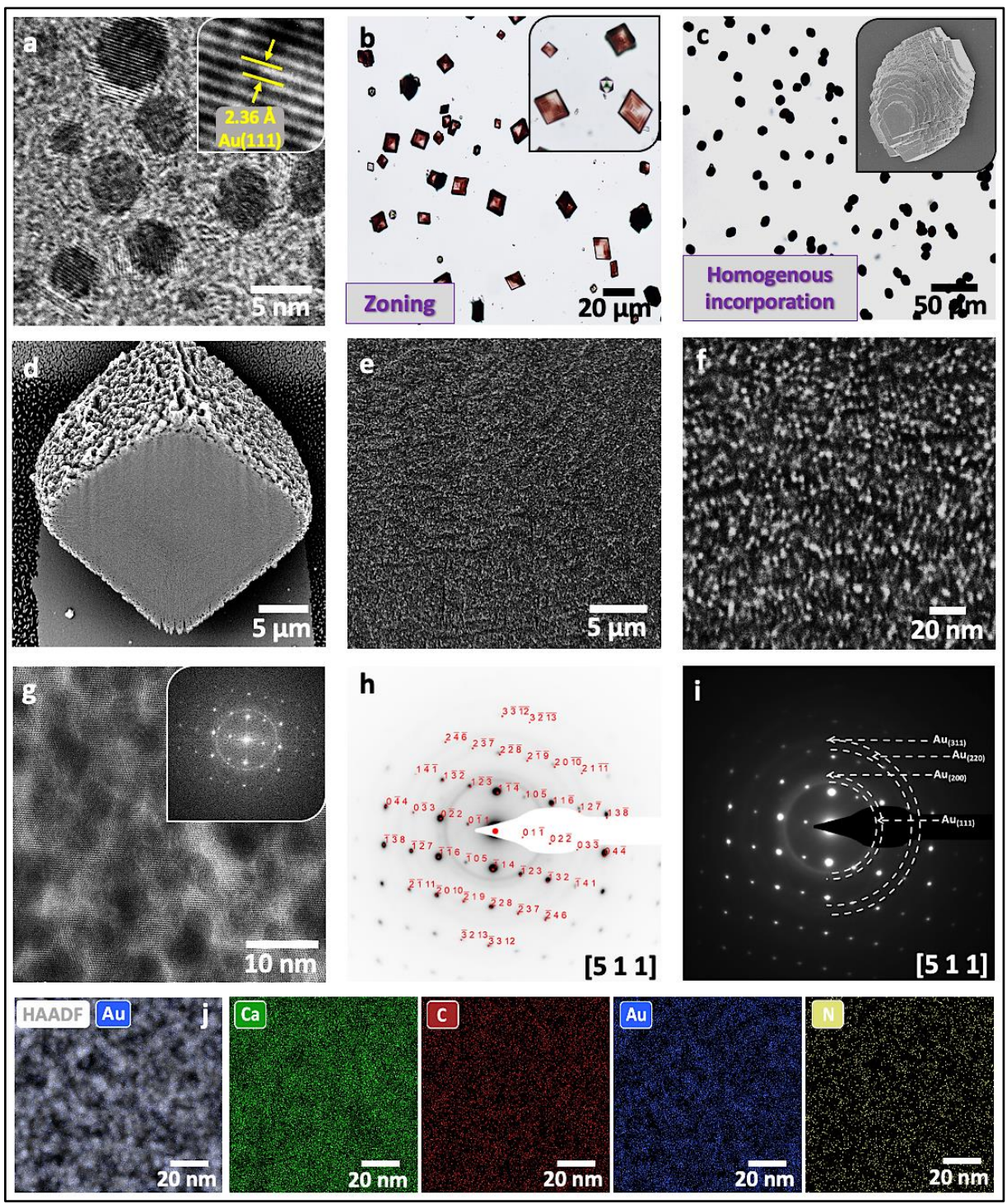

**Figure 6.** (a) TEM image of ≈ 5 nm AuNPs functionalized with branched PEI (Au/PEI). The inset is a HRTEM image showing Au (111) lattice fringes. (b) Optical image of calcite crystals precipitated at $[Ca^{2+}]$ = 10 mM and [Au/PEI] = 0.01 wt%, showing zoning effects. (c) Optical micrograph and SEM image (inset) of dark and elongated calcite crystals precipitated at $[Ca^{2+}]$ = 10 mM and [Au/PEI] = 0.1 wt%. (d-f) FIB-SEM images through crystals revealing the dense

and uniform incorporation of the AuNPs (bright spots) in calcite. (g) HRTEM and corresponding FFT of a cross-section through a nanocomposite crystal, displaying lattice fringes that demonstrate the single crystallinity of the calcite host. (h-i) SAED patterns of the nanocomposites showing diffraction spots of calcite single crystals and rings corresponding to Au. (j) HAADF-STEM and EDX-STEM maps showing the uniform distribution of Ca, C (*i.e.,* calcite) and Au, N (*i.e.,* Au/PEI nanoparticles) throughout the hybrid crystals.

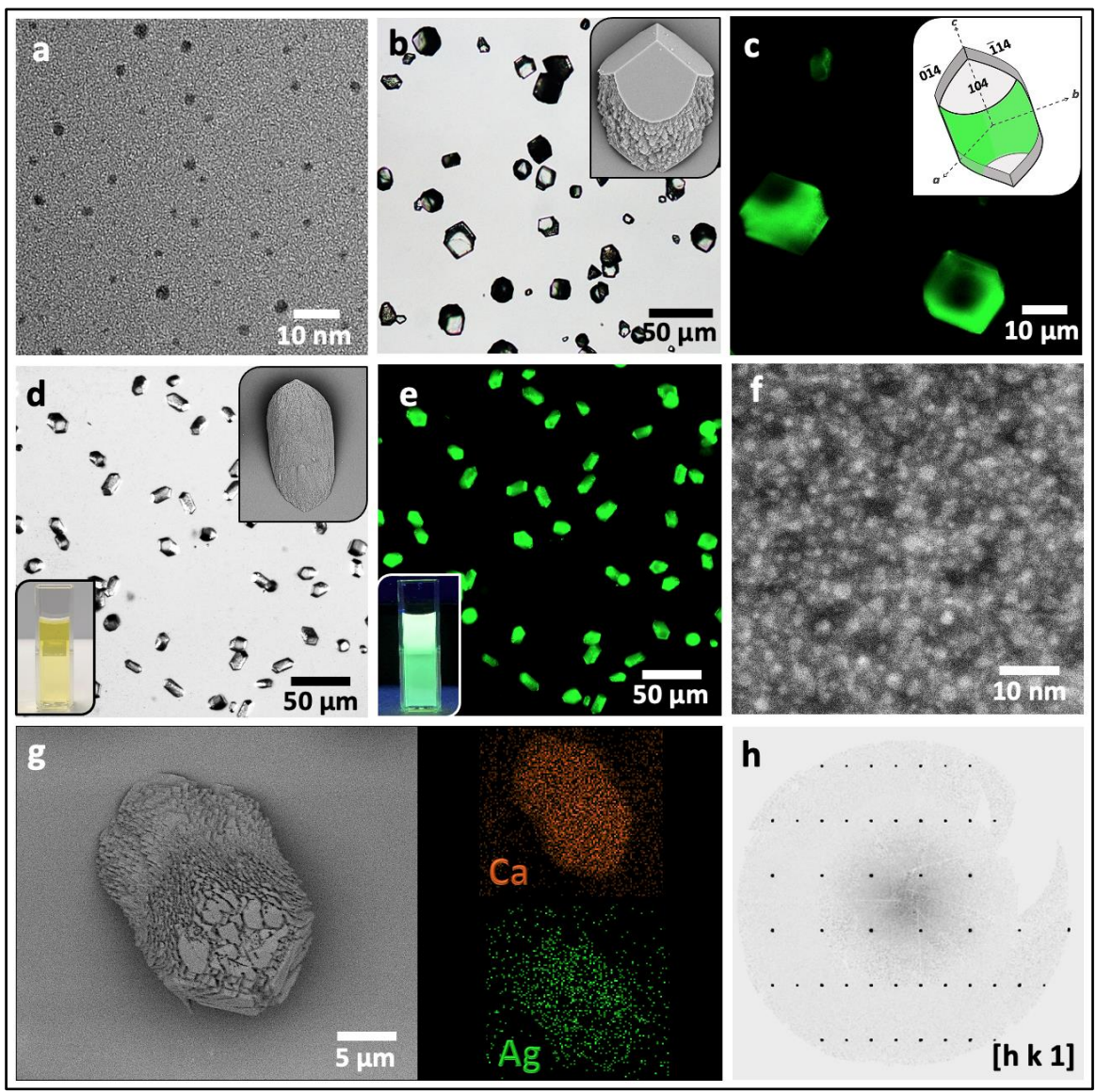

**Figure 7.** (a) TEM micrograph of Ag nanoclusters functionalized with branched PEI (Ag/PEI). (b) Optical and SEM (inset) images of calcite precipitated at $[Ca^{2+}]$ = 10 mM and [Ag/PEI] = 0.01 wt% and (c) corresponding fluorescence image of the hybrid crystals showing zoning. (d) Optical image, SEM image (top right) and digital image (bottom left) of calcite precipitated at $[Ca^{2+}]$ = 10 mM and [Ag/PEI] = 0.1 wt%, and corresponding (e) fluorescence and (f) dark-field TEM images demonstrating homogeneous incorporation of Ag/PEI. (g) SEM-EDX maps showing elemental Ca and Ag of the composite crystals precipitated at $[Ca^{2+}]$ = 10 mM and [Ag/PEI] = 0.1 wt%, and (h) corresponding XRD pattern demonstrating the single crystal character of the hybrid crystals.

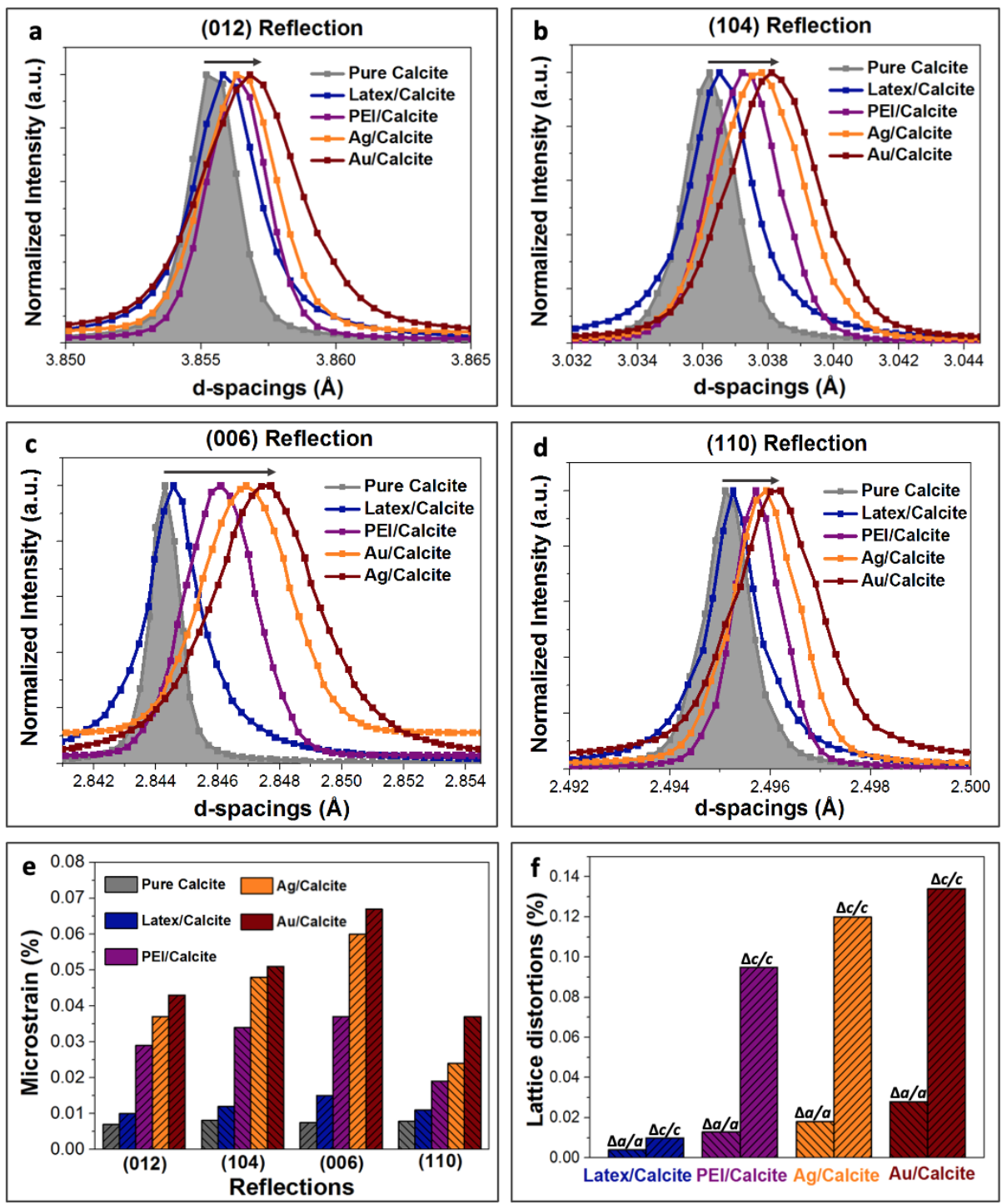

**Figure 8.** (a-d) HRPXRD patterns and (e) microstrain fluctuations of the indicated reflections of pure calcite (grey), PMMA-PEI latex/calcite (blue), PEI/calcite (purple), Ag/calcite (orange) and Au/calcite (red). (f) Lattice distortions along the *a*-axis and *c*-axis of calcite incorporating the diverse cationic additives.

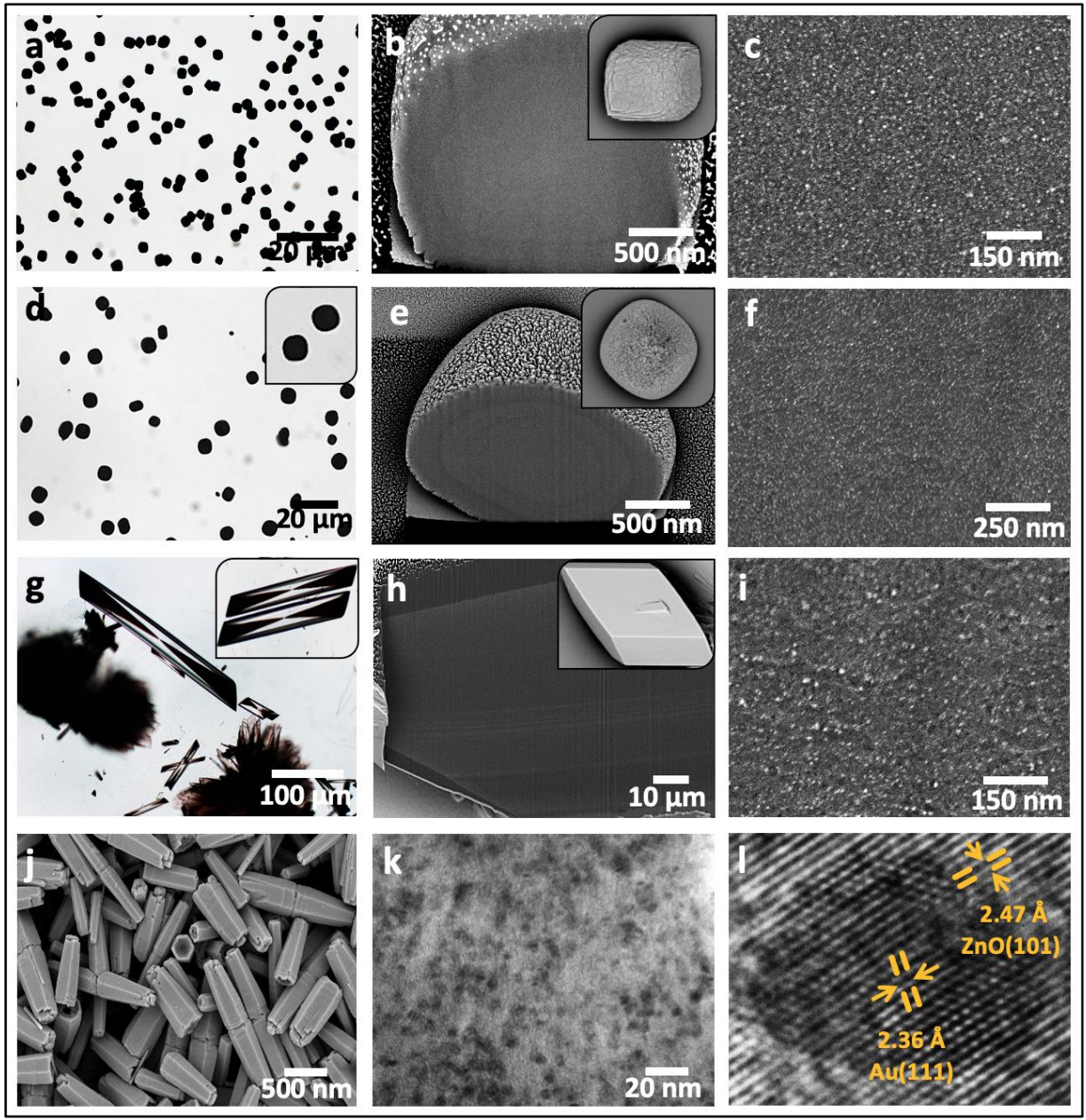

**Figure 9.** (a-c) MnCO$_3$, (d-f) SrSO$_4$, (g-i) CaSO$_4\cdot$2H$_2$O and (j-l) ZnO nanocomposite crystals incorporating Au/PEI nanoparticles. (a, d and g) are optical images, while (j) shows a SEM image of the Au/ZnO nanocrystals. (b-c, e-f, and h-i) FIB-SEM cross-section images revealing the high levels of AuNP incorporation in the various inorganic host crystals. (k) TEM micrograph of a thin section through an Au/ZnO nanocomposite showing the uniformly incorporated nanoparticles, (l) and corresponding HRTEM image displaying the lattice fringes of Au (111) and ZnO (101).

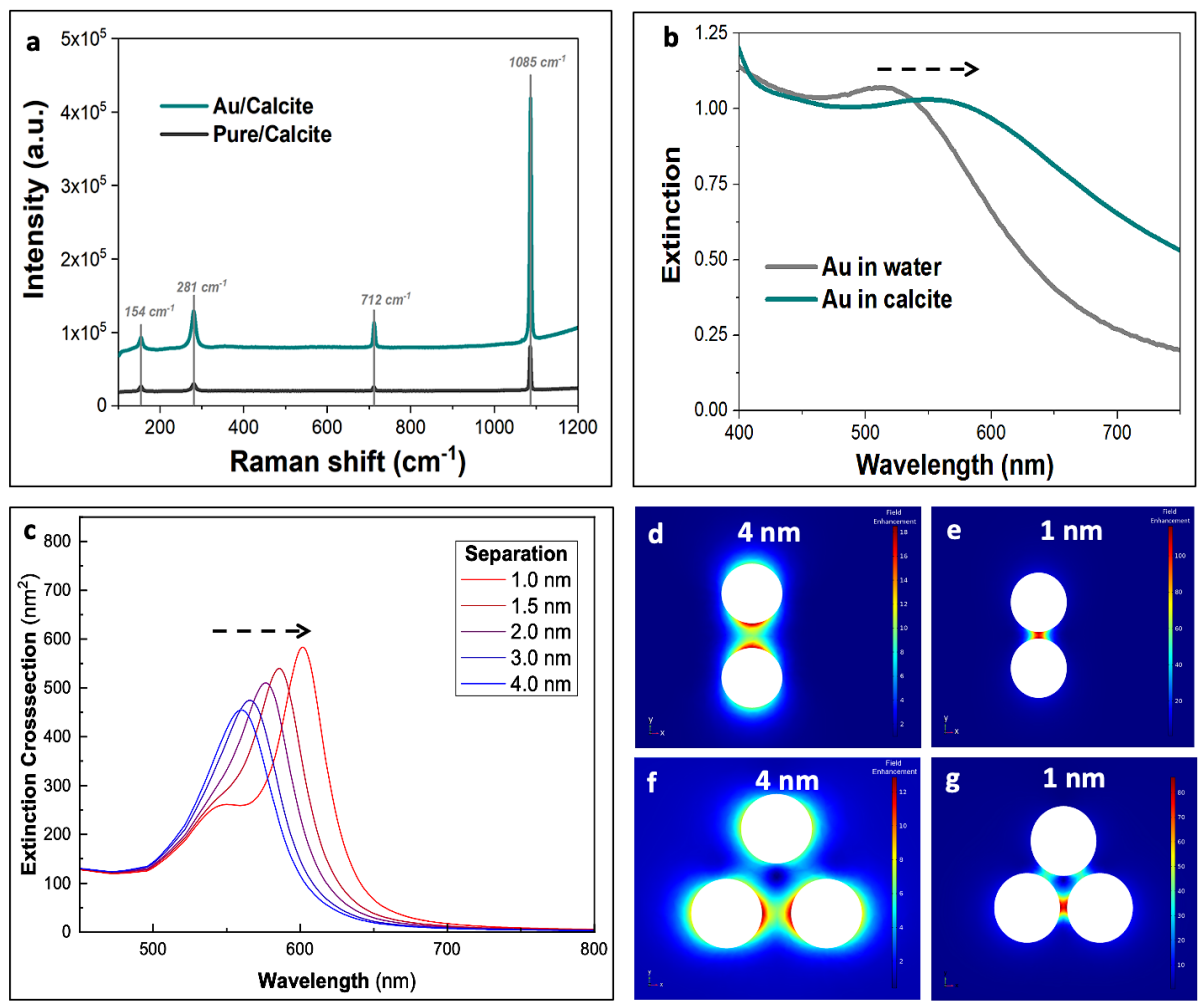

**Figure 10.** (a) Raman spectra of pure crystals (grey) and Au/calcite nanocomposites (cyan), showing a surface-enhanced Raman scattering (SERS) effect for the composites when irradiated with a laser source ($\lambda$ = 785 nm), using 10 sec exposure time and 5% laser power. The spectra are offset for clarity. (b) UV-visible extinction spectra of Au/PEI in pure water (grey) and in calcite (cyan), displaying a red-shift of the surface plasmon resonance band of the AuNPs when incorporated in calcite. (c) Simulated extinction spectra for 5 nm AuNPs dimers occluded in calcite with varying the interparticle separation between 1 nm and 4 nm. (d-g) FEM simulated field enhancement $|E|/|E_0|$ along the y-axis for the 5 nm AuNPs dimers (d-e) and trimers (f-g) with an interparticle gap of 4 nm (d and f) and 1 nm (e and g).